\documentclass[a4paper,11pt]{article}
\pdfoutput=1 

\usepackage{jcappub} 
\usepackage[T1]{fontenc} 
\usepackage{amsmath}
\usepackage{color}

\def\bk{{\bf{k}}}

\def\bk{{\bf k}}
\def\bq{{\bf q}}

\def\k{{\bf k}}

\title{\boldmath Adding helicity to inflationary magnetogenesis}

\author[a]{Chiara Caprini}
\author[a,b,c]{Lorenzo Sorbo}

\affiliation[a]{CNRS, URA 2306 and CEA, IPhT, F-91191 Gif-sur-Yvette, France}
\affiliation[b]{APC (Astroparticules et Cosmologie), UMR 7164 (CNRS, Universit\'e Paris 7)\\ 10, rue Alice Domon et L\'eonie Duquet, 75205 Paris Cedex 13, France}
\affiliation[c]{Department of Physics, University of Massachusetts, Amherst, MA 01003, USA}

\emailAdd{chiara.caprini@cea.fr}
\emailAdd{sorbo@physics.umass.edu}

\abstract{The most studied mechanism of inflationary magnetogenesis relies on the time-dependence of the coefficient of the gauge kinetic term $F_{\mu\nu}\,{F}^{\mu\nu}$. Unfortunately, only extremely finely tuned versions of the model can consistently generate the cosmological magnetic fields required by observations. We propose a generalization of this model, where also the pseudoscalar invariant $F_{\mu\nu}\,\tilde{F}^{\mu\nu}$ is multiplied by a time dependent function. The new parity violating term allows more freedom in tuning the amplitude of the field at the end of inflation. Moreover, it leads to a helical magnetic field that is amplified at large scales by magnetohydrodynamical processes during the radiation dominated epoch. As a consequence, our model can satisfy the observational lower bounds on fields in the intergalactic medium, while providing a seed for the galactic dynamo, if inflation occurs at an energy scale ranging from $10^5$ to $10^{10}$~GeV. Such energy scale is well below that suggested by the recent BICEP2 result, if the latter is due to primordial tensor modes. However, the gauge field is a source of tensors during inflation and generates a spectrum of gravitational waves that can give a sizable tensor to scalar ratio $r={\cal O}(0.2)$ even if inflation occurs at low energies. This system therefore evades the Lyth bound. For smaller values of $r$, lower values of the inflationary energy scale are required. The model predicts fully helical cosmological magnetic fields and a chiral spectrum of primordial gravitational waves.
}

\begin{document}
\maketitle
\flushbottom

\section{Introduction}

The origin of cosmological magnetic fields is still mysterious. While an astrophysical origin is not excluded, the lower bound of $\sim 10^{-18}\div 10^{-15}$G recently inferred~\cite{Neronov:1900zz,Taylor:2011bn,Vovk:2011aa} for fields with a correlation length larger than $\sim 1$~Mpc has rekindled the interest in the construction of inflationary mechanisms of magnetogenesis. 

Conformal invariance of standard electromagnetism implies that an expanding flat Universe does not  generate any magnetic field.  Therefore one has to go beyond standard electrodynamics~\cite{Turner:1987bw}. Since mechanisms that violate gauge invariance generally give rise to ghost-like instabilities \cite{Himmetoglu:2008zp,Himmetoglu:2009qi}, most of the attention has gone to two mechanisms that preserve gauge invariance. The most commonly studied case is the one where the Lagrangian for the gauge field reads $-\frac{I^2}{4}\,F_{\mu\nu}\,F^{\mu\nu}$, with $I$ a function of time: this has been first analyzed in details by Ratra in~\cite{Ratra:1991bn} (for conciseness we will therefore refer to this model  as the ``Ratra model''). Another possibility is the axion model~\cite{Garretson:1992vt}, where the Lagrangian for the gauge field reads $-\frac{1}{4}\,F_{\mu\nu}\,F^{\mu\nu}-\frac{\phi}{4\,f}\,F_{\mu\nu}\,\tilde{F}^{\mu\nu}$, with $\phi$ a rolling pseudoscalar  and with the constant $f$ measuring the strength of the scalar-photon coupling. 

In the Ratra model, the spectral index of the magnetic field is controlled by the behavior of the function $I$, so that a scale invariant magnetic field can be obtained by assuming that $I\propto a^n$, where $a$ is the scale factor of the Universe during inflation, provided $n$ is appropriately chosen to either $n=-3$ or $n=2$. A scale invariant magnetic field and an inflationary energy scale of the order of the GUT scale would lead to cosmological fields of $\sim 10^{-12}$~G. Unfortunately, such choices of $n$ are associated either to an excessive energy~\cite{Martin:2007ue} in the electric field ($n=-3$) or to a strongly coupled theory~\cite{Demozzi:2009fu} during inflation ($n=2$). For $-2<n\le 0$ both the requirement of controllable energy in the electric field and that of weak coupling throughout inflation can be satisfied. However this part of the parameter space leads to a blue spectrum that cannot yield magnetic fields stronger than $10^{-32}$~G~\cite{Demozzi:2009fu} at $1$~Mpc today, even in the best case scenario of inflation occurring at the highest energy scale, of the order of $10^{16}$~GeV, allowed by observations\footnote{Stronger magnetic fields can be obtained by choosing $n<-2$ and assuming an infrared cutoff associated to the beginning of magnetogenesis during inflation~\cite{Martin:2007ue,Ferreira:2013sqa}. In this case, fields in agreement with observations can be achieved if inflation takes place at $\sim 10$~MeV.}.

Axion magnetogenesis is characterized by a very blue spectrum $B(k)\propto k^2$ that would make it phenomenologically irrelevant~\cite{Garretson:1992vt}. However, in reference~\cite{Anber:2006xt} it was observed that, being produced by a parity violating background, the magnetic field is maximally helical. As several works~\cite{inverse,Banerjee:2004df} have shown, helical fields can undergo, during the radiation and matter dominated eras, a process of inverse cascade that transfers power from small to large scales. Moreover, in the case of the Ratra model, the initial amplitude and correlation length of the magnetic field are both determined by the Hubble scale $H$ during inflation, whereas in axion magnetogenesis those two quantities are controlled by two independent parameters, $H$ and the strength of the coupling. Therefore, in axion magnetogenesis one can lower the energy of inflation, and correspondingly increase the initial correlation length of the magnetic field, without having to reduce at the same time the overall magnitude of the magnetic field. As a consequence, the constraints of \cite{Martin:2007ue,Fujita:2012rb,Fujita:2014sna} do not apply to the axion model, nor to the model presented in the following. In reference~\cite{Anber:2006xt} it was argued that in the axion model magnetic fields of $\sim 10^{-23}$G at $1$ Mpc could be generated by suitably lowering the energy scale of inflation to $\sim 10^{12}$GeV. As we will see, the system presented here includes the model of~\cite{Garretson:1992vt,Anber:2006xt} as a special case. Our analysis will therefore update the one of~\cite{Anber:2006xt} as well as the subsequent one of \cite{Durrer:2010mq}.

In the present paper we describe a hybrid of the Ratra and of the axion model. We will consider a $U(1)$ gauge field whose Lagrangian has the form
\begin{equation}\label{lagr}
{\cal L}=I^2(\tau)\,\left(-\frac{1}{4}F_{\mu\nu}\,F^{\mu\nu}+\frac{\gamma}{8}\epsilon_{\mu\nu\rho\lambda}F^{\mu\nu}\,F^{\rho\lambda}\right)\,,
\end{equation}
with $\gamma$ a constant that will turn out to be of ${\cal O}(10)$. The overall function $I$ can be chosen in such a way that both strong coupling and backreaction by the electric field are avoided. This implies that, even if the field will not be scale invariant, it can have a spectral tilt that is less blue than the one obtained in pure axion magnetogenesis.  The presence of the parameter $\gamma$ leads to a magnetic field with a net helicity, so that inverse cascade is at work. 

We will consider the implications of the model~(\ref{lagr}) for {\em (i)} providing the ``seed'' fields assumed to be the origin, via a dynamo mechanism, of the observed galactic fields (see \cite{Brandenburg:2004jv} and references therein), and {\em (ii)} fulfilling the bound on cosmological fields originating by the non-observation of cascade photons in the GeV band from TeV blazars~\cite{Neronov:1900zz,Taylor:2011bn,Vovk:2011aa}. We will find that both observational bounds can be satisfied in our model, when the energy scale of inflation is of the order of $10^5\div 10^{10}$~GeV.

One important constraint on the model originates from the requirement that the inhomogeneous magnetic field does not mix too strongly with the fluctuations of the inflaton, leading to a nongaussian component in the metric perturbations that would exceed the limits from Planck~\cite{Barnaby:2010vf,Barnaby:2011vw}. A  way out of this problem is to assume that the field associated to the time dependence of the function $I$ is not the inflaton, but some other field~\cite{Barnaby:2012xt}. Once the constraints from~\cite{Barnaby:2010vf,Barnaby:2011vw} are evaded this way, the main effect on metric perturbations of the gauge field is to contribute to the spectrum of primordial tensor modes~\cite{Barnaby:2012xt,Cook:2013xea,Shiraishi:2013kxa}. In the present work we will assume that the $B$-mode polarization observed by the BICEP2 experiment~\cite{Ade:2014xna} is entirely of primordial origin, and we will therefore set the tensor-to-scalar ratio $r=0.2$ (it is straightforward to generalize our analysis to arbitrary values of $r$, c.f. figure~\ref{fig:hubnv}). As stated above, for our model to give interesting results inflation should occur below $\sim 10^{11}$~GeV. As a consequence, the contribution to $r$ by the standard amplification of the vacuum fluctuations of the metric is completely negligible: this model violates the Lyth bound~\cite{Lyth:1996im} in a way that is analogous to the one discussed first in~\cite{Sorbo:2011rz} and subsequently in~\cite{Cook:2011hg,Barnaby:2012xt,Mukohyama:2014gba}, since it displays a large tensor-to-scalar ratio without Planckian excursions of the inflaton.

Our model comes with two specific signatures. First, the current magnetic fields should be helical (see for instance~\cite{Tashiro:2013ita} for prospects of a direct measure of the helicity of cosmological magnetic fields). Moreover, since the gauge field has a definite helicity, the tensor modes produced during inflation will be parity-odd, and as a consequence will generate non-vanishing $\langle TB\rangle$ and $\langle EB\rangle$ in the Cosmic Microwave Background (CMB) radiation~\cite{Lue:1998mq,Caprini:2003vc}. For primordial tensor modes with $r={\cal O}(0.1)$, a satellite mission should be able to determine whether $\langle TB\rangle$ and $\langle EB\rangle$ are actually non vanishing~\cite{Saito:2007kt,Contaldi:2008yz,Gluscevic:2010vv,Ferte:2014gja}.

The plan of the paper is the following. In section 2 we study the amplification of the vacuum fluctuations  of gauge fields due to the Lagrangian~(\ref{lagr}).  In section 3 we study the production of gravitational waves by the electromagnetic field during inflation. In section 4 we study the subsequent evolution of the magnetic field after the end of inflation, and analyze how the inverse cascade increases power in the magnetic modes at large scales. In section 5 we discuss the parts of parameter space that lead to magnetic fields in agreement with observations. In section 6 we present two simple models, derived from supergravity and from brane inflation, where the Lagrangian~(\ref{lagr}) is obtained. In section 7 we discuss our results and we conclude.

\section{Production of electromagnetic fields during inflation}
\label{sec:production}

We start from the Lagrangian~(\ref{lagr}), where we assume  $I(\tau)=a^n(\tau)=(-H\,\tau)^{-n}$,  with $a$ the scale factor of the Universe and $a_{\rm end}=1$ its value at the end of inflation. We choose $n\le 0$ to have weak coupling throughout inflation and $n>-2$ to avoid infrared divergences in the energy in the electric field (see section 3). Also, we choose $\gamma>0$ without loss of generality. 

In the Coulomb gauge $A_0=\partial^i\,A_i=0$, upon integration by parts, and defining $I\,A=\tilde{A}$ for the canonically normalized field, the Lagrangian becomes
\begin{equation}
{\cal L}=\frac{1}{2}\,\tilde{A}_i'{}^2-\frac{1}{2}\left(\nabla \tilde{A}_i\right)^2+\frac{1}{2}\,\frac{I''}{I}\,\tilde{A}_i^2-\gamma\,\frac{I'}{I}\,\epsilon_{ijk}\,\tilde{A}_i\,\partial_j\,\tilde{A}_k\,.
\end{equation}
Quantization of the gauge field on helicity modes $\tilde{A}_\sigma$ with $\sigma=\pm 1$ is realized by defining
\begin{equation}\label{helicity}
\tilde{A}_i(\bk)=\sum_{\sigma=\pm} \int\frac{d^3\bk}{(2\pi)^{3/2}}\,\epsilon_i^\sigma(\k)\,e^{i\bk\,{\bf x}}
\left(\tilde{A}_\sigma(\bk,\,\tau)\,\hat{a}_\sigma(\bk)+\tilde{A}^*_\sigma(-\bk,\,\tau)\,\hat{a}^\dagger_\sigma(-\bk)\right)\,
\end{equation}
where $\epsilon_i^\sigma(\k)$ is the helicity$-\sigma$ vector and where $\tilde{A}_\sigma$ satisfies
\begin{equation}\label{eqasigma}
\tilde{A}_\sigma''+\left(k^2+2\,\sigma\,\xi\,\frac{k}{\tau}-\frac{n\,\left(n+1\right)}{\tau^2}\right)\,\tilde{A}_\sigma=0\,,
\end{equation}
where we have defined\footnote{In the following, we will consider also the case where $\xi$ is finite and $n=0$, even if this formally implies $\gamma\to\infty$, as it covers the special case of purely axionic magnetogenesis of~\cite{Garretson:1992vt,Anber:2006xt}.} $\xi\equiv -n\,\gamma$. Guided by the results found in previous analyses (see for example reference~\cite{Anber:2006xt}) we will assume $\xi\gg 1$, which enhances the amplitude of the gauge field. We will see in the following that the range of phenomenologically relevant values for this parameter is $\xi={\cal O}(10)$, c.f. the discussion after figure~\ref{fig:hubnv}. 

Eq.~(\ref{eqasigma}) shows that the system evolves in three stages. At early times the ultraviolet term $k^2$ dominates and the photons are in their Bunch-Davies vacuum. Later on, when the term proportional to $\xi$ dominates, the overall sign of the coefficient of $\tilde{A}_\sigma$  in eq.~(\ref{eqasigma}) is determined by the sign of $\sigma\,\xi$: the mode functions for which  $\sigma\,\xi>0$ are exponentially amplified (remember $\tau<0$), whereas the photons of opposite helicity do not feel such an amplification. At this stage a net chirality in the photon system is generated. Finally, as $\tau\to 0$, mode functions of both helicities are amplified by the term $-n\,\left(n+1\right)/\tau^2$, and the spectral index at large scales is controlled by the parameter $n$. The final result of this process is a field with an arbitrary spectral index and with net helicity.

Since modes for which $\sigma\,\xi<0$ are less amplified we will neglect their effect altogether, assuming $\sigma=+1$ in what follows. The explicit solution of eq.~(\ref{eqasigma})  that goes to positive-frequency only modes at $\tau\to-\infty$ is a linear combination of the Coulomb wave functions:
\begin{align}\label{asigma}
\tilde{A}_\sigma(k,\,\tau)=\frac{1}{\sqrt{2\,k}}\,&\left(G_{-n-1}(\xi,-k\,\tau)+i\,F_{-n-1}(\xi,-k\,\tau)\right).
\end{align}

For $|k\,\tau|\ll \xi$, i.e., for the phenomenologically interesting scales corresponding to modes that are out of the Bunch-Davies vacuum, the contribution of $F_{-n-1}(\xi,\,-k\,\tau)$ to $\tilde{A}_\sigma(k,\,\tau)$ is negligible. Moreover, using the result of~\cite{Durrer:2010mq}, we can approximate in the same regime
\begin{align}\label{aapprox}
\tilde{A}_\sigma(k,\,\tau)\simeq \sqrt{-\frac{2\,\tau}{\pi}}\,e^{\pi\xi}\,K_{-2n-1}\left(\sqrt{-8\,\xi\,k\,\tau}\right)\,,
\qquad\qquad |k\,\tau|\ll \xi\,,\,\, \xi\gg 1\,.
\end{align}
In particular, for $|k\,\tau|\ll 1/\xi$ we obtain
\begin{align}\label{aapproxsm}
\tilde{A}_\sigma(k,\,\tau)\simeq \sqrt{-\frac{\tau}{2\,\pi}}\,e^{\pi\xi}\,\Gamma\left(|2n+1|\right)\,\left|2\,\xi\,k\,\tau\right|^{-|n+1/2|}\,,\qquad\qquad |k\,\tau|\ll 1/\xi\ll 1\,,
\end{align}
that shows that the overall amplitude of the magnetic field is exponentially large in $\xi$. 

Defining the magnetic field power spectrum as $k^3\langle B_i({\bf k})B^*_i({\bf q}) \rangle = 2(2\pi)^3 P_B(k)\delta({\bf k} - {\bf q})$, with $B_i({\bf k})=k A_i({\bf k})$, eq.~(\ref{aapproxsm}) also shows that the spectral index of the magnetic field $\sqrt{P_B(k)}\propto k^{n_B}$ at large scales is 
\begin{equation}\label{nb}
n_B=\frac{5}{2}-\left|n+\frac{1}{2}\right|\,.
\end{equation}
We have used this definition of the spectral index since it corresponds to the scaling with $k$ of the magnetic field intensity on a given scale $\ell=2\pi/k$, c.f. also eq.~(\ref{bell}). We have therefore that $n=-3$ corresponds to a scale invariant magnetic field. This would however imply a red electric field, whose energy would quickly dominate the system invalidating our analysis~\cite{Martin:2007ue}. The case $n=-2$ corresponds to a flat electric spectrum, still leading to a logarithmic infrared divergence, that  in its turn would lead, among other effects, to a breaking of the $SO(3)$ invariance of our inflating patch~\cite{Bartolo:2012sd}, and to anisotropic expansion \cite{Kanno:2009ei}. For this reason we will focus on the regime $n>-2$ where all relevant quantities are infrared-finite. We remind also that $n\leq 0$ to avoid strong coupling \cite{Demozzi:2009fu}.

Before studying the intensity of the magnetic field at cosmological scales, however, let us focus on the effects of the presence of the gauge field during inflation and on the corresponding constraints on the space of parameters of the model.

\section{Electromagnetic fields, low scale inflation and the Lyth bound}%

Typically, an upper bound on the intensity of the magnetic field during inflation is imposed by requiring that the metric perturbations induced by the gauge field do not affect the properties of the CMB. If $I$ is a function of the inflaton, then strong constraints emerge from the requirement that nongaussianities are under control~\cite{Barnaby:2010vf,Barnaby:2011vw,Nurmi:2013gpa}. In the present work we will assume, along the lines of~\cite{Barnaby:2012xt}, that  the inflaton is not directly coupled to the gauge field and the function $I$ is associated to some other field that is rolling during inflation. In the case of pure axion magnetogenesis the strongest constraint originates from the requirement that the gauge modes do not overproduce gravitational waves~\cite{Barnaby:2012xt}, even if additional  interesting signatures might appear in the three-point correlators in the CMB~\cite{Cook:2013xea,Shiraishi:2013kxa}. We expect the situation to be similar in our model. As a consequence, in this section we will compute the spectrum of tensor modes induced by a magnetic field present during inflation (for an analysis of scalar modes, see e.g. \cite{Bonvin:2011dt,Bonvin:2013tba}). This calculation is a straightforward extension of that presented in~\cite{Sorbo:2011rz}, that was valid in the specific case $n=0$, $\xi\neq 0$. The equation of motion for the helicity-$\lambda$ tensor mode $h_\lambda$ reads
\begin{equation}\label{eqh}
h_\lambda''+2\,\frac{a'}{a}\,h_\lambda'+k^2\,h_\lambda=\frac{2}{M_P^2}\,\Pi_{\lambda}{}^{ij}(\bk)\,T^{EM}_{ij}(\bk)
\end{equation}
where we have introduced the polarization tensors $\Pi_\lambda^{ij}({\bk})=\frac{1}{\sqrt{2}}\epsilon_{-\lambda}^i({\bk})\,\epsilon_{-\lambda}^j({\bk})$. In eq.~(\ref{eqh}) $T^M_{ij}$ represents the spatial part of the stress-energy tensor of the gauge field $T^M_{ij}=-I^2\,a^{-2}\,A_i'\,A_j'+\dots$, where $\dots$ stands for the contribution proportional to $\delta_{ij}$, that is projected out by $\Pi_{\lambda}{}^{ij}$, and for the part proportional to spatial derivatives of $A_i$, that can be neglected as the magnetic field gives negligible contribution when $\xi\gtrsim {\cal O}(1)$~\cite{Sorbo:2011rz}. Note that after reheating the conductivity of the universe becomes effectively infinite: the electric field, which is the dominant component during inflation, is therefore quickly dissipated and only the magnetic field remains in the radiation era. For an analysis of the dissipation effect after inflation, see e.g. \cite{Martin:2007ue}. 

We can now promote the functions $h_\pm$ to operators $\hat{h}_\pm$. Neglecting for the time being the solution of the homogeneous part of eq.~(\ref{eqh}), the expression of ${\hat h}_\pm$  reads
\begin{align}
{\hat h}_\pm(\bk)=-\frac{2}{M_P^2}\int d\tau'\,G_k(\tau,\,\tau')\,(-H\,\tau')^{2-2\,n}\int\frac{d^3{\bq}}{(2\pi)^{3/2}}\,\Pi_\pm^{lm}({\bk})\,\hat{A}_l'({\bq},\tau')\,\hat{A}_m'({\bk}-{\bq},\tau')\,.
\end{align}
where $G_k(\tau,\tau')$ is the retarded Green function for the operator $\frac{d^2}{d\tau^2}-\frac{2}{\tau}\,\frac{d}{d\tau}+k^2$:
\begin{align}
G_k(\tau,\,\tau')=\frac{1}{k^3\,\tau'{}^2}\left[\left(1+k^2\,\tau\,\tau'\right)\sin k\left(\tau-\tau'\right)+ k \left(\tau'-\tau\right)\,\cos k\left(\tau-\tau'\right)\right]\,\Theta(\tau-\tau')\,, 
\end{align}
where $\Theta$ is the Heaviside function.

Using the decomposition~(\ref{helicity}) for the photon, and neglecting the contribution from the negative helicity mode, we obtain, for the two-point function of the tensor fluctuations at the end of inflation $\tau\simeq 0$,
\begin{align}\label{longint}
\langle h_\lambda({\bk})\,h_\lambda({\bk}')\rangle&=\frac{H^{4-4\,n}}{32\,\pi^3\,M_P^4}\,\delta(\bk+\bk')
\int d^3\bq \left(1+\lambda\frac{\bk\cdot\bq}{k\,q}\right)^2\,\left(1+\lambda\frac{(\bk-\bq)\cdot\bk}{|\bk-\bq|\,k}\right)^2\nonumber\\
&\times\left|\int d\tau'\,\tau'^{2-2\,n}\,G_k(0,\,\tau')A_+'(\bq,\,\tau')\,A_+'(\bk-\bq,\,\tau')\right|^2\,,
\end{align}
where the second line depends on the helicity of the gravitons and has been obtained using the following property of the helicity projectors
\begin{equation}
\left|\epsilon^i_{\lambda}({\bf p}_1)\,\epsilon^i_{\lambda'}({\bf p}_2)\right|^2=\frac{1}{4}\left(1-\lambda\lambda'\,\frac{{\bf p}_1\cdot {\bf p}_2}{p_1\,p_2}\right)^2\,.
\end{equation}

\begin{figure}[tbp]
\centering
\includegraphics[width=10.cm,trim=0cm 0cm 0cm
  0cm,clip]{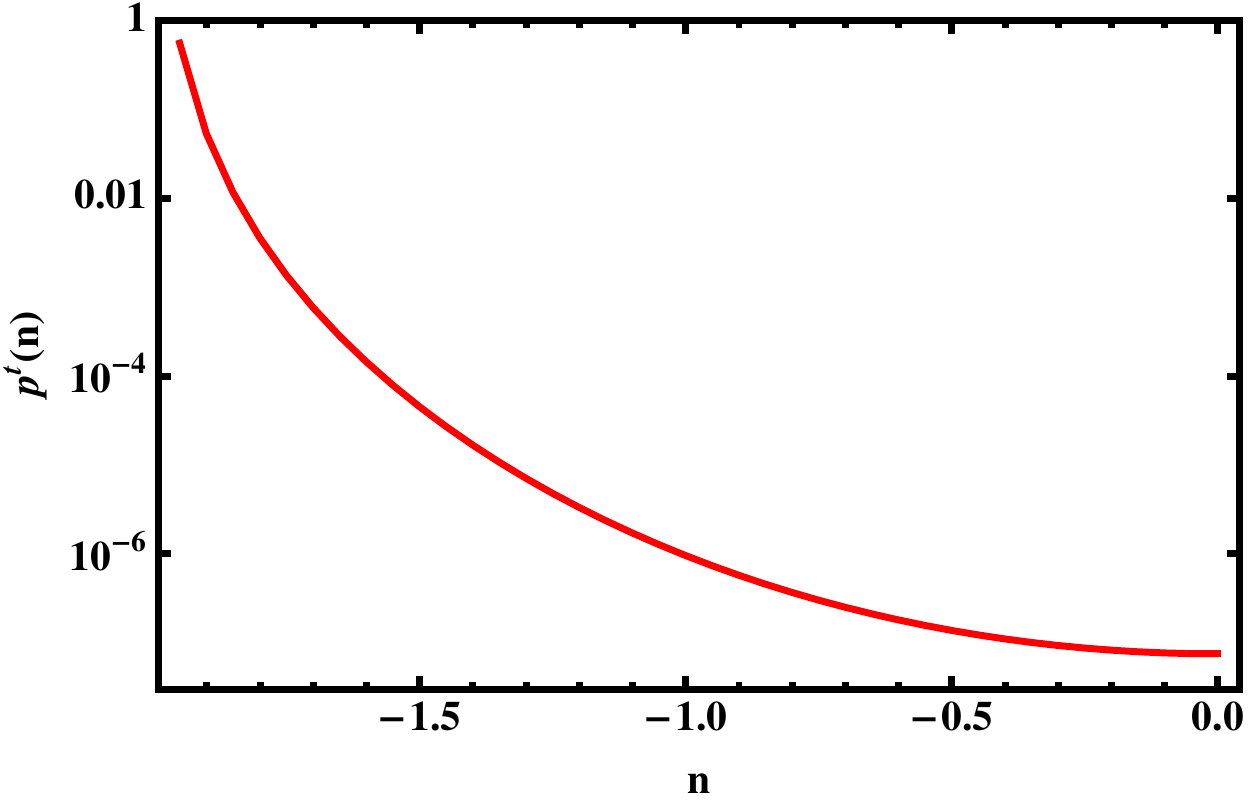}
  \caption{The function $p^t(n)$ appearing in equation~(\ref{ptem}) that determines the amplitude of the tensor spectrum induced by the magnetic field. 
    }\label{fig:ftn}
\end{figure}

To compute the expression~(\ref{longint}) we use the approximate form~(\ref{aapprox}) of the mode functions $\tilde{A}_+\equiv I\,A_+$. We also approximate $G_k(0,\,\tau')\simeq \tau'/3$, since, due to the exponential suppression of $A_+(k,\,\tau')$ for $|k\,\tau|\gg \xi^{-1}$, we can assume $|k\,\tau'|\ll 1$. Finally, we consider only the contribution with $\lambda=+1$, as negative helicity gravitons are produced much less abundantly. The integral in $d\tau'$ can then be performed analytically, whereas the integral in $d^3{\bf q}$ has to be computed numerically. We thus obtain the following scale invariant power spectrum for tensors
\begin{equation}\label{ptem}
{\cal P}^t= p^t(n)\,\frac{H^4}{M_P^4}\,\frac{e^{4\,\pi\,\xi}}{\xi^{6}}\,
\end{equation}
where the function $p^t(n)$ is plotted in figure~\ref{fig:ftn} for $-1.95\leq n\leq 0$. 

We note that the function $p^t(n)$ diverges as $n\to -2$, as for $n=-2$ the energy of the electric field is logarithmically divergent in the infrared. For this reason our analysis is limited to $n>-2$. In analogy to what was done in~\cite{Martin:2007ue,Demozzi:2009fu,Ferreira:2013sqa}, it would be possible to extend our analysis also to the regime $n<-2$ at the cost of introducing a new parameter in the model corresponding to an infrared cutoff for the gauge field (which is in one-to-one correspondence with the time of the beginning of magnetogenesis during inflation). While we will not pursue this option here, it would be interesting to see how the constraints that we will find for the model~(\ref{lagr}) would be relaxed by such an assumption.

Expressing the amplitude of the tensors generated by the gauge field during inflation in terms of the tensor-to-scalar ratio $r$ we obtain the relation
\begin{equation}\label{lyth}
\frac{H}{M_P}\,\frac{e^{\pi\,\xi}}{\xi^{3/2}}=\left(\frac{r\,{\cal P}_\zeta}{p^t(n)}\right)^{1/4}\,,
\end{equation}
where ${\cal P}_\zeta\simeq 2.5\times 10^{-9}$ is the amplitude of the scalar perturbations. 

In the present paper, to fix ideas, we will assume that the $B$-modes observed by BICEP2 are entirely of primordial origin and we will therefore set $r=0.2$. This will allow to reduce the number of free parameters of the model from $3$ ($n,\,\xi,\,H$) to $2$. If the constraint $r=0.2$ is relaxed to an upper bound, then eq.~(\ref{lyth}) will turn into an upper bound for $H$ once $n$ and $\xi$ are fixed (c.f. the dashed line in Fig.~\ref{fig:hubnv}). 

As we will see, the generation of a sufficiently strong magnetic field requires a Hubble parameter several orders of magnitude below the one required by a tensor-to-scalar ratio $r=0.2$ if the tensor modes are the result of amplification of vacuum fluctuations by the de Sitter geometry. Therefore the contribution to ${\cal P}^t$ from tensors produced by the standard mechanism -- that is, those associated to the solution of the homogeneous component of eq.~(\ref{eqh}) -- is largely subdominant with respect to that given by eq.~(\ref{ptem}). Since in this system one can obtain low scale inflation while obtaining a large value of $r$, it provides a counterexample to the Lyth bound, along the lines of~\cite{Sorbo:2011rz,Cook:2011hg,Barnaby:2012xt,Mukohyama:2014gba} (see also~\cite{Senatore:2011sp} for related ideas).

\section{Inverse cascade and the current intensity of the magnetic field}%

In order to evaluate the amplitude and the correlation scale of the magnetic field today, here we study its time evolution after the end of inflation. For simplicity, we will assume instantaneous reheating. Several numerical and analytical studies~\cite{inverse,Banerjee:2004df} show that helical magnetic fields undergo a process of inverse cascade during the radiation dominated epoch. During this process, the comoving correlation scale of the magnetic field increases and its comoving intensity decreases, and power is transferred from small to large scales while the magnetic spectrum at scales larger than the correlation scale maintains its spectral index unchanged, displaying a property of self-similarity. 

In agreement with~\cite{Durrer:2013pga}, we define the correlation scale as
\begin{equation}
L=\frac{\int d^3{\bk}\,\frac{2\pi}{k}\,\left|B({\bf k})\right|^2}{\int d^3{\bk}\,\left|B({\bf k})\right|^2}
\end{equation}
where, for the helical fields under consideration, $B({\bf k})\equiv k\,A_+({\bf k})$, with $A_+({\bf k})$ given by the approximate expression~(\ref{aapprox}). In particular, at the end of inflation
\begin{equation}\label{lrh}
L_{\rm {rh}}=\frac{18\,\pi}{(3-2\,n)\,(5+2\,n)}\,\frac{\xi}{H}\,\,.
\end{equation}
We also define the intensity $B$ of the magnetic field as 
\begin{equation}
B^2\equiv \langle B^2\rangle=\int\frac{d^3\k}{(2\pi)^3}\,\left|k\,A_+\right|^2\,,
\end{equation}
so that, at the end of inflation
\begin{equation}\label{brh}
B_{\rm {rh}}^2=H^4\,\frac{e^{2\,\pi\,\xi}}{\xi^5}\,\frac{\Gamma(4-2\,n)\,\Gamma(6+2\,n)}{2^{8}\times 3^2\times 5\times7\times\pi^3}\,.
\end{equation}

Starting from these initial conditions we can compute the present  values $B_0$ and $L_0$ of the magnetic field and of its correlation length. The analysis of reference~\cite{Banerjee:2004df} (see also \cite{Durrer:2013pga} for a compendium) shows that the evolution of $B$ and $L$ during the post-inflationary era goes through several alternating turbulent, viscous, and free-streaming phases. The occurrences of these phases depend on the values of $B_{\rm {rh}}$ and $L_{\rm {rh}}$ and on the evolution with temperature of the kinetic viscosity of the plasma (which is determined by the particle species with the longest mean free path: neutrinos, followed by photons after neutrino decoupling). 

For a large set of initial conditions the system starts in a turbulent phase, and the inverse cascade can take place: the comoving value of $B$ decreases and that of $L$ increases. As the mean free path of neutrinos grows, the system enters a viscous phase, the magnetic field decouples from the fluid and the comoving values of $B$ and $L$ stay constant. When the neutrino mean free path grows beyond the typical scale of the flow $L$, the free-streaming phase begins and $B$ decreases again while $L$ grows. After neutrino decoupling, the same evolution pattern repeats with photons. For all reasonable initial conditions given by our model, the system never goes back to the turbulent phase, and the flow is in the free-streaming phase at recombination. 

\begin{figure}[tbp]
\centering
\includegraphics[width=10.cm,trim=0cm 0cm 0cm
  0cm,clip]{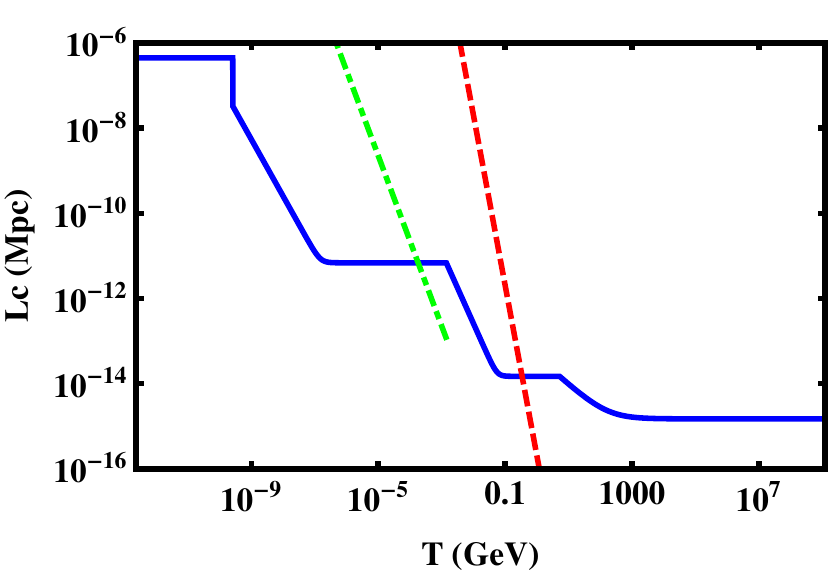}
  \caption{The evolution of the comoving magnetic correlation scale as a function of temperature, for $n=-1.9$ and $\xi=13$, corresponding to an inflationary energy scale $\rho_{\rm {inf}}^{1/4}\simeq 3\times 10^9$~GeV. The red, dashed curve represents the comoving neutrino mean free path, and the green, dot-dashed curve the photon one. The jump in $L_c(T)$ at recombination is due to the sudden increase of the photon mean free path just before decoupling, caused by the drop in the ionization fraction. The alternating phases of turbulent, viscous, and free-streaming evolution are apparent. The evolution of $B$ is obtained straightforwardly from that of $L$ by imposing conservation of helicity.
    }\label{fig:turbulence}
\end{figure}

Further non-trivial evolution in the matter era occurs only for very weak values of $B$, and the turbulent cascade in the matter era resumes to almost constant comoving values for $B$ and $L$: we therefore conservatively assume that comoving $B$ and $L$ always stay constant after recombination. The evolution of $L$ can be followed in figure \ref{fig:turbulence}. 

As a matter of fact, the details of the time evolution are not very relevant, as long as the system experiences turbulent evolution. Reference~\cite{Banerjee:2004df} demonstrates that the properties of the flows are such that the evolution of the plasma fluid at any moment is described by the simple causality relation 
\begin{equation}\label{mhd}
v_L=H\,L\,,
\end{equation}
where $H$ is the Hubble parameter and $v_L$ the fluid velocity on the flow scale $L$. When turbulence is active the system is in equipartition, so that $v_L\simeq v_A\equiv B/\sqrt{\rho+p}$ ,where $\rho$ and $p$ are the energy density and pressure of the fluid particles that couple to the magnetic field; in all other cases, $v_L\ll v_A$. In particular, in the free-streaming photon phase right before recombination, with photon mean free path $\ell_{\rm mfp}^{\gamma}$, one has that $v_L\simeq v_A^2\, \ell_{\rm mfp}^{\gamma} / L$~\cite{Banerjee:2004df}, and the above relation (\ref{mhd}) gives $v_A\simeq L\sqrt{H_{\rm rec}/\ell_{\rm mfp}^{\gamma}}\simeq L\, H_{\rm rec}$. The last equality holds since the photon mean free path gets comparable to the Hubble scale at recombination. It appears therefore that the system satisfies the same causality relation as if it was in a turbulent regime: the details of the time evolution do not matter much. 

The relation $v_A\simeq L \,H_{\rm rec}$ links the values of $B$ and $L$ at recombination. Further evolving until today under the condition that their comoving values stay constant, one finds that the following relation holds~\cite{Banerjee:2004df,Durrer:2013pga}: 
\begin{equation}\label{blprop}
B_0\simeq 10^{-8}\,{\rm G}\,\left(\frac{L_0}{\rm {Mpc}}\right)\,.
\end{equation}
As stated in references~\cite{Banerjee:2004df,Durrer:2013pga}, this relation is very general and works for a large set of initial conditions. It can also be obtained by imposing that the system is turbulent in the late Universe (which is certainly a reasonable assumption after reionisation), and applying eq.~(\ref{mhd}) with $v=v_A$ today: this reads $B_0/\sqrt{\rho_b^0}\simeq H_0\, L_0$, with $\rho_b^0$ the baryon density. The difference among the values of $B_0$ found by evolving (\ref{mhd}) evaluated at recombination until today, as opposed to assuming its validity today, is of the order (we set $\rho_{\rm cdm}\simeq \rho_{\rm rec}/2$ at recombination) of $(\sqrt{\Omega_b/2})/\Omega_{\rm cdm}$: this can be considered of order one within the level of precision with which eq.~(\ref{mhd}) holds. 

In order to determine the values of $B_0$ and $L_0$ we need to supplement eq.~(\ref{blprop}) with a second relation. Since our magnetic field is maximally helical, such a relation is provided by the conservation of (comoving) helicity, applying during the regimes of high conductivity (i.e., always during the expansion of the Universe). Helicity is defined as the volume integral
\begin{equation}
{\cal H}=\int d^3{\bf x}\,{\bf A}\cdot{\bf B}\propto B^2\,L\,,
\end{equation}
so that, factoring in the expansion of the Universe, conservation of helicity implies
\begin{equation}\label{helcon}
B_0^2\,L_0=B_{\rm {rh}}^2\,L_{\rm {rh}}\,\left(\frac{a_{\rm {rh}}}{a_0}\right)^3\,.
\end{equation}
Combining eqs.~(\ref{blprop}) and~(\ref{helcon}) we determine the current values of the magnetic field and the correlation scale. Note that the spectral index of the magnetic field is given by eq.~(\ref{nb}): therefore, at scales $\ell>L$, the intensity of the magnetic field is
\begin{equation}\label{bell}
B(\ell)=B\left(\frac{L}{\ell}\right)^{\left(5-|2\,n+1|\right)/2}\,.
\end{equation}
We can then insert the obtained values of the magnetic field and the correlation scale into eq.~(\ref{bell}) to get the predicted value of the magnetic field at a given scale $\ell$. 

\section{Observational constraints}%
%
We consider two classes of observations that can lead to constraints on primordial magnetic fields:

\begin{enumerate}

\item magnetic fields in Galaxies, of the order of $10^{-6}$~G, are observed with a number of techniques. They are typically assumed to be the end product of the mean-field dynamo mechanism, able to amplify a weak seed field by several orders of magnitude (for a review, see~\cite{Brandenburg:2004jv}). There are large uncertainties on the amplitude of the seed field required to explain the observed value, depending on the details of complicated galactic magnetohydrodynamics. Typical numbers are in the $10^{-23}\div 10^{-21}$~G range at Mpc comoving scales (also accounting for the amplification due to the collapse of the galaxy). Note that the validity of the mean-field dynamo is questioned by the observation of microGauss level magnetic fields in protogalactic clouds at high redshift \cite{Kronberg:2007dy,Bernet:2008qp}; however, we will still consider it here as providing a reference value for the seed amplitude;

\item more recently, the non-observation, by the Fermi telescope, of cascade photons in the GeV band from TeV blazars and active galactic nuclei~\cite{Neronov:1900zz} has led to a lower bound on the intensity of magnetic fields in the intergalactic medium (IGM) of the order of $6\times 10^{-18}\div 10^{-16}$~G~\cite{Vovk:2011aa} for fields with correlation lengths of $1$~Mpc or more. This bound is strengthened by a factor $\sqrt{1\,{\mathrm {Mpc}}/L_0}$ if the correlation length $L_0$ is smaller than $1$~Mpc.

\end{enumerate}

We note that the first class of observations leads to a constraint on the intensity of the magnetic field, irrespective of its correlation length, at a given length scale; the second, on the other hand, is a constraint on the magnetic field and on its correlation length.

It is questionable whether primordial magnetic fields are effectively necessary to initiate the galactic dynamo and account for the magnetic fields observed in galaxies. Astrophysical generation mechanisms based on charge separation (such as the Biermann battery), or ejection of magnetic fields from stars have also been invoked as possible seeds for the dynamo (for reviews, see e.g. \cite{vallee,Kulsrud:2007an}). 

On the other hand, it would be difficult for astrophysical generation mechanisms or for generation mechanisms related to structure formation to provide fields that can account for the lower bound on the magnetic field amplitude in the IGM: the reason being, that this bound applies in the voids among matter structures, in the absence of high density matter or ionized plasma \cite{Dolag}. 

Therefore, the second class of observations can be considered as a stronger hint on the necessity of primordial magnetic fields. As a consequence, we first focus on the part of parameter space of our model able to satisfy the requirements from this latter type of observational constraint. Once we have identified the values of the parameters for which our generation mechanism can satisfy the lower bound in the IGM, we proceed to evaluate the strength of the magnetic field at the Mpc scale and investigate whether it can account for the seed field of the dynamo.  

Note that a magnetic field with small initial amplitude of the order of $10^{-23}\div 10^{-21}$~G on the Mpc scale cannot explain the observations of magnetic fields in clusters by simple amplification by adiabatic collapse. Clusters magnetic fields have an amplitude of about $10^{-6}$~G and appear to be correlated at the cluster scale  \cite{Govoni:2004as}, necessitating a seed field of the order of $10^{-9}$~G on the Mpc scale. On the other hand, it is not excluded that such fields could be explained simply by ejection of galactic magnetic fields, which are then processed by small scale turbulence in the intra-cluster medium and amplified up to the microGauss level \cite{Donnert:2008sn}. Therefore, primordial magnetic fields that satisfy the lower bound in the IGM and act as a seed for the galactic dynamo could, at least in principle, provide an explanation for the magnetization of matter structures. 

\begin{figure}[tbc]
\centering
\includegraphics[width=10.cm,trim=0cm 0cm 0cm
  0cm,clip]{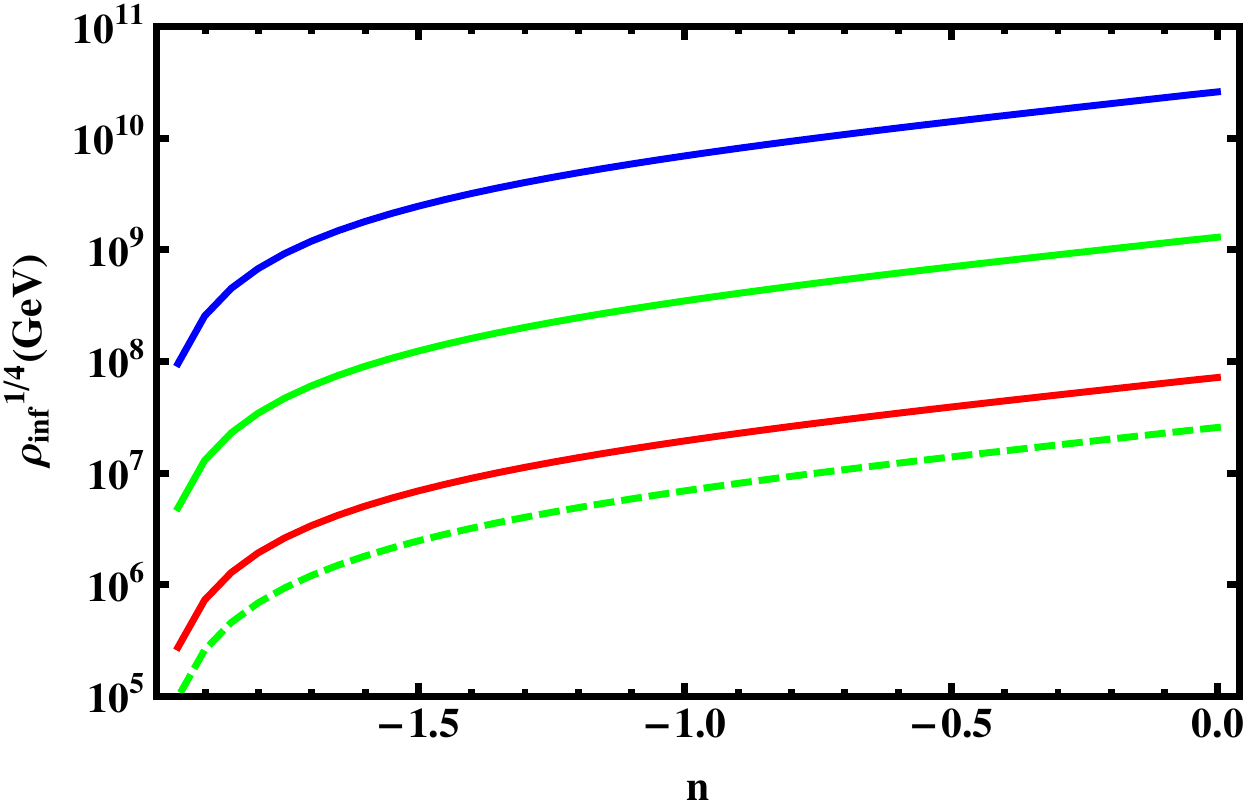}
  \caption{Solid, top to bottom: inflationary energy scale, as a function of the parameter $n$, necessary for the generation of a magnetic field of $6\times 10^{-18}$~G (the lower bound quoted by~\cite{Vovk:2011aa}), $2.5\times 10^{-17}$~G and $10^{-16}$~G  times  $\sqrt{1\,{\mathrm {Mpc}}/L_0}$, assuming a tensor-to-scalar ratio $r=0.2$. Dashed line: inflationary energy scale, as a function of the parameter $n$, necessary for the generation of a magnetic field of $2.5\times 10^{-17}$~G at scales of $1$~Mpc, assuming a tensor-to-scalar ratio $r=10^{-4}.$
    }
  \label{fig:hubnv}
\end{figure}

We plot in figure~\ref{fig:hubnv} the value of $\rho_{\rm {inf}}^{1/4}\equiv (3\,M_P^2\,H^2)^{1/4}$ necessary to yield magnetic fields of $6\times 10^{-18}$~G (the lower bound quoted in~\cite{Vovk:2011aa}), $2.5\times 10^{-17}$~G and $10^{-16}$~G times  $\sqrt{1\,{\mathrm {Mpc}}/L_0}$ for $-1.95<n<0$, assuming a tensor-to-scalar ratio $r=0.2$. Note that we multiply by the factor $\sqrt{1\,{\mathrm {Mpc}}/L_0}$ to account for the IGM lower bound relevant at the correlation scale $L_0$, given by our generation mechanism.

Figure~\ref{fig:hubnv} is the main result of the present paper. It shows that even in the worst case we considered, that is for $n=-1.95$, magnetic fields of the order of $10^{-16}\,{\rm G}\times \sqrt{1\,{\mathrm {Mpc}}/L_0}$ can be generated if the inflationary scale is of the order of $10^5$~GeV. In the best case, that is for $n=0$, fields of the same intensity can be obtained for inflation occurring at $10^8$~GeV. Magnetic fields matching $6\times 10^{-18}$~G times $\sqrt{1\,{\mathrm {Mpc}}/L_0}$, the lower bound quoted by~\cite{Vovk:2011aa}, can be obtained, for $n=0$, if inflation occurs at $\sim 10^{10}$~GeV. We also plot in the same figure the  energy scale of inflation required for the generation of a field of $2.5\times 10^{-17}\,{\rm G}\times \sqrt{1\,{\mathrm {Mpc}}/L_0}$ in our model, under the assumption that the tensor to scalar ratio is $r=10^{-4}$. 

The values of the coupling parameter $\xi$ in figure~\ref{fig:hubnv} increase with decreasing $n$ and with increasing magnetic field strength, and are comprised between $12<\xi<19$ if $r=0.2$. In the case $r=10^{-4}$, slightly larger values of $\xi$ are necessary to account for a magnetic field with intensity $2.5\times 10^{-17}\,{\rm G}\times \sqrt{1\,{\mathrm {Mpc}}/L_0}$ with respect to the case $r=0.2$: as $n$ varies, we find $14<\xi<17$ if $r=0.2$ and correspondingly $16<\xi<19$ if $r=10^{-4}$. At any rate, we obtain that the range of phenomenologically relevant values for this parameter is $\xi={\cal O}(10)$, as anticipated in section \ref{sec:production}.

Since the constraints from~\cite{Vovk:2011aa} do not depend on the spectral index of the magnetic field, whereas the constraints on $n$ from the production of gravitational waves get stronger as $n$ decreases towards $-2$ (see figure~\ref{fig:ftn}), the curves in figure~\ref{fig:hubnv} are increasing functions of $n$. In particular, the curves giving $\rho_{\rm {inf}}^{1/4}$ as a function of $n$ go to zero as $n\to -2$, since for $n=-2$ the amplitude of the spectrum of tensors induced by the magnetic field is formally divergent. 

\begin{figure}[tbc]
\centering
\includegraphics[width=10.cm,trim=0cm 0cm 0cm
  0cm,clip]{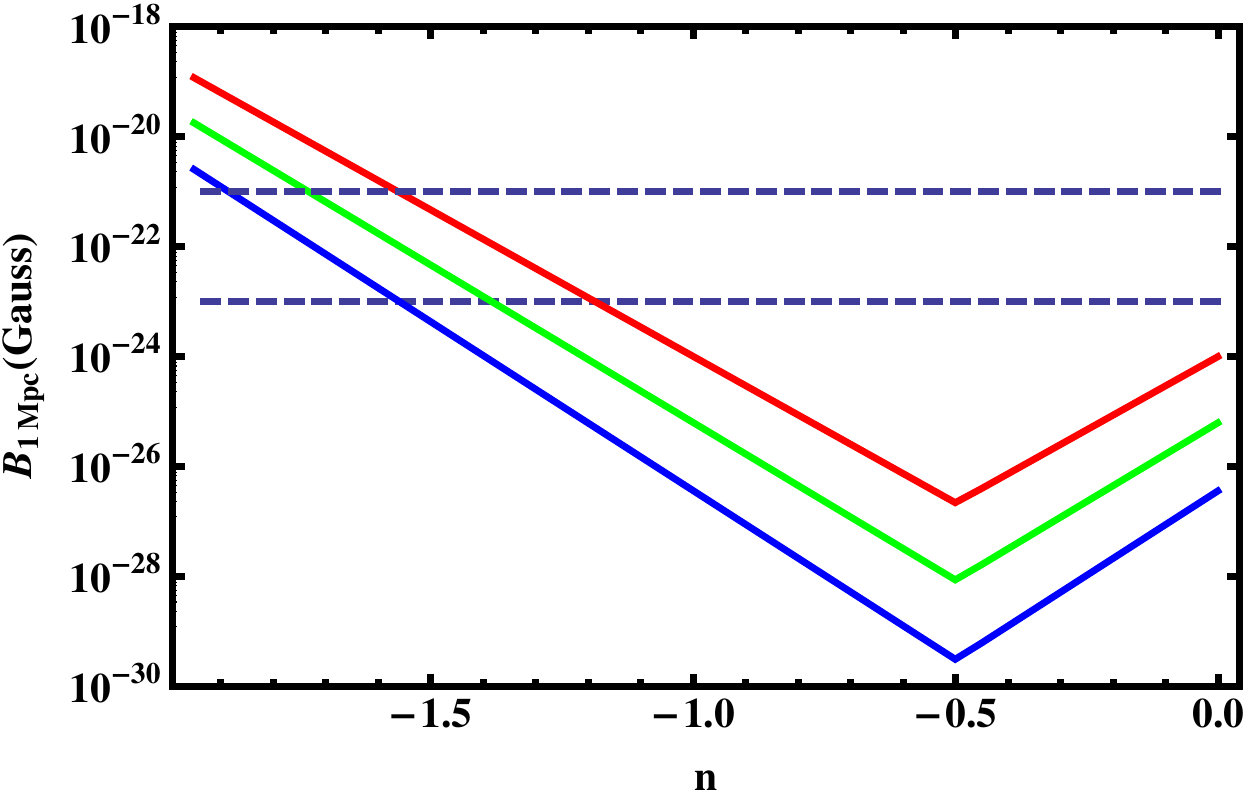}
  \caption{Solid lines: intensity of the magnetic field at $1$~Mpc scale, as a function of the parameter $n$, obtained by imposing that the magnetic field  at the correlation scale $L_0$ satisfies the observational lower bound in the IGM: from top to bottom, $6\times 10^{-18}$~G (the lower bound quoted by~\cite{Vovk:2011aa}), $2.5\times 10^{-17}$~G and $10^{-16}$~G  times  $\sqrt{1\,{\mathrm {Mpc}}/L_0}$. As in figure \ref{fig:hubnv}, we  assume a tensor-to-scalar ratio $r=0.2$. The kink at $n=-0.5$ originates from the absolute value in the spectral index, eq.~(\ref{nb}). The two horizontal dashed lines correspond to the weaker ($10^{-23}$~G) and stronger ($10^{-21}$~G) requirements for the field intensities able to seed the galactic dynamo.
    }\label{fig:fighb1mpc}
\end{figure}

Once we choose the parameters of the model in such a way to agree with the IGM lower bound~\cite{Neronov:1900zz,Vovk:2011aa}, we can check whether the constraints for the galactic dynamo are also satisfied by reading out of eq.~(\ref{bell}) the intensity of the magnetic fields at the Mpc scale. We plot such a quantity in figure~\ref{fig:fighb1mpc}, for the same choices of parameters considered in figure~\ref{fig:hubnv}. Comparison of figures~\ref{fig:hubnv} and~\ref{fig:fighb1mpc} shows that, even if for $n=0$ the IGM lower bound is satisfied, the magnetic field at the Mpc scale is too weak for the dynamo ($\sim 10^{-24}$~G for the choice of parameters leading to $10^{-16}$~G$\times\sqrt{1\,{\mathrm {Mpc}}/L_0}$), and is therefore a poor candidate for providing the seed for galactic magnetic fields. On the other hand, for the values of the parameters leading to the same intensity $10^{-16}$~G$\times\sqrt{1\,{\mathrm {Mpc}}/L_0}$, a redder magnetic field spectrum with $n\simeq -1.9$ provides a higher magnetic field amplitude on the Mpc scale, of the order of $10^{-19}$~G, which is a valuable seed for the magnetic fields observed in galaxies. This comes at the price of a lower energy scale of inflation, which is nevertheless still well above the TeV scale.

Note that the magnetic fields generated in this model also satisfy the constraints derived in \cite{Caprini:2009pr} from the production of gravitational waves in the radiation era from reheating until Nucleosynthesis. 
\\

\section{Two simple models: supergravity and brane inflation}%

In this section we briefly describe two models that lead to the effective Lagrangian~(\ref{lagr}). The first one is derived in the context of $N=1$ four-dimensional supergravity. We start from the general kinetic term for a gauge field~(see e.g.~\cite{Nilles:1983ge})
\begin{equation}
{\cal L}=-\frac{1}{4}\,{\rm Re}\left\{f\right\}\,F_{\mu\nu}\,F^{\mu\nu}-\frac{1}{4}\,{\rm Im}\left\{f\right\}\,F_{\mu\nu}\,\tilde{F}^{\mu\nu}
\end{equation}
where the gauge kinetic function $f$ is a holomorphic function of the superfields in the theory. We then assume that $f$ is given by the product of two fields $f(X,\,Y)=X\,Y$ and we require that both the real and the imaginary part of $Y$ as well as the imaginary part of $X$ are stabilized to 
\begin{align}
{\rm {Re}}\{Y\}=Y_0\,,\quad {\rm {Im}}\{Y\}=\gamma\,Y_0\,,\quad {\rm {Im}}\{X\}=0\,,
\end{align}
so that the Lagrangian for the gauge field reads, defining $X_R\equiv {\rm {Re}}\{X\}$,
\begin{equation}
{\cal L}=X_R\,Y_0\,\left(-\frac{1}{4}\,F_{\mu\nu}\,F^{\mu\nu}-\frac{\gamma}{4}\,F_{\mu\nu}\,\tilde{F}^{\mu\nu}\right)
\end{equation}
that, identifying $I^2\equiv X_R\,Y_0$, reproduces our Lagrangian~(\ref{lagr}).

Another example of a model leading to a generalization of the Lagrangian~(\ref{lagr}) was discussed, in the context of brane inflation, in reference~\cite{Langlois:2009ej} -- see eq.~(83) in that paper, to be confronted with our eq.~(\ref{eqasigma}). In \cite{Langlois:2009ej} the dilaton, that determines the coefficient of the term $F_{\mu\nu}\,F^{\mu\nu}$, differs from the axion, that multiplies $F_{\mu\nu}\,\tilde{F}^{\mu\nu}$. A different scaling with time of these two coefficients would lead to a broken spectrum for the gauge field. The analysis of such a system is beyond the scope of the present work, but it might lead to interesting phenomenology.

\section{Conclusions}%

Magnetogenesis during inflation is appealing since it has the advantage, with respect to causal generation mechanisms, of being able to produce magnetic fields with interesting amplitudes at large scales. However, it has been demonstrated that the most commonly studied gauge invariant model of magnetogenesis is affected by the backreaction and the strong coupling constraints: the constraints apply right in the region of parameter space for which the model could produce magnetic fields with observationally relevant intensity \cite{Martin:2007ue,Demozzi:2009fu}. 

Some solutions have been proposed, for example in~\cite{Martin:2007ue,Ferreira:2013sqa,Demozzi:2012wh}: however, in order to produce high enough magnetic fields, these models require a low value for the energy scale of inflation. They are therefore somehow unnatural, especially if the recent BICEP2~\cite{Ade:2014xna} measurement will prove to be due to primordial tensor modes~\cite{Fujita:2012rb,Ferreira:2014hma}. 

In this paper we have presented a model for magnetogenesis that satisfies the backreaction and the strong coupling constraints, and can both account for the observational lower bound on the magnetic field amplitude in the IGM~\cite{Neronov:1900zz,Taylor:2011bn,Vovk:2011aa}, and provide high enough magnetic seeds to explain, through the galactic dynamo, the microGauss fields observed in galaxies. The energy scale of inflation depends on the parameters of the model, but it is in any case higher than about $2\times 10^5$ GeV. This is not in contrast with the level of tensor-to-scalar ratio that would be implied by a primordial origin of the BICEP2 signal, since the gauge field responsible for the magnetic field today acts also as a source of tensor modes, generating a scale invariant spectrum of gravitational waves whose amplitude depends on the parameters of the model. Standard gravitational waves due to the amplification of vacuum fluctuations of the background metric are largely subdominant with respect to those directly sourced by the gauge field, because of the low energy scale at which inflation is occurring: therefore, the model presented here evades the Lyth bound~\cite{Lyth:1996im}. 

Production of gravitational waves during inflation turns out to induce the strongest constraints on the parameters of the model here proposed. Non-gaussianity constraints on the spectrum of scalar perturbations generated by the gauge field could also be of importance~\cite{Barnaby:2010vf,Barnaby:2011vw}, but we avoid these latter by assuming that the scalar field directly coupled to the gauge field is not the inflaton. 

We impose that the level of tensor modes generated by the gauge field does not overcome $r=0.2$. Furthermore, we impose that the magnetic field amplitude at the correlation scale today (which is always smaller than 1 Mpc) satisfies the lower bound in the IGM, both in its most conservative form and in two more stringent ones. From these conditions, we obtain the maximal allowed value of the energy scale of inflation as a function of the spectral index of the gauge field power spectrum. We then calculate the amplitude of the resulting magnetic field today on the Mpc scale, and find that it is sufficiently high to seed the galactic dynamo, provided that the magnetic field power spectrum is shallow enough. 

The success of the proposed mechanism resides on the combination of a Ratra-like coupling~\cite{Ratra:1991bn} with an axion-like coupling \cite{Garretson:1992vt}. The Ratra-like coupling allows to control the spectral index of the magnetic field power spectrum, rendering it less steep than in the case of the pure axion-like coupling: therefore, the amplitude of the magnetic field on large (Mpc) scales can be higher than what obtained in this latter. On the other hand, thanks to the axion-like coupling, the intensity of the magnetic field depends exponentially on the strength of the coupling. Consequently, the magnetic field intensity can be enhanced, despite the fact that we are always in the regime of weak coupling for the Ratra-like term, $n\leq 0$. Moreover, the presence of the axion-like coupling implies parity violation, so that the resulting magnetic field is maximally helical. The field can therefore undergo a process of inverse cascade during its time evolution in the radiation dominated era, with transfer of magnetic field energy at scales larger than the typical correlation scale. The combination of these three effects results in the necessary energy scale of inflation being higher than in previously proposed models, rendering this model more natural from this point of view.

\acknowledgments

We thank Rajeev Kumar Jain, Andrii Neronov, S\'ebastien Renaux-Petel and Martin Sloth for very useful discussions. The work of L.S. is partially supported by the U.S. National Science Foundation grant PHY-1205986.

\end{document}